\documentclass[conference]{IEEEtran}

\usepackage{listings}
\usepackage{color}
\usepackage{subcaption}
\usepackage{algpseudocode}
\usepackage{booktabs}

\usepackage{url}
\usepackage{amsmath}
\usepackage{graphicx} 
\usepackage[colorlinks,linkcolor=red,citecolor=blue,urlcolor=blue,draft]{hyperref}
\usepackage[flushleft]{threeparttable}
\usepackage{algorithm}
\usepackage{etoolbox}
\usepackage{tabularx}
\usepackage{hhline}
\usepackage{mdwlist}
\usepackage{multirow}
\usepackage{rotating}

\definecolor{dkgreen}{rgb}{0,0.6,0}
\definecolor{dkorange}{rgb}{0.7,0.2,0}
\input{Definitions}

\iftrue
\newcommand{\sji}[1]{{\color{blue}[\textbf{\sc Shihao}: \textit{#1}]}}
\newcommand{\satish}[1]{{\color{green}[\textbf{\sc Satish}: \textit{#1}]}}
\newcommand{\sli}[1]{{\color{red}[\textbf{\sc Sheng}: \textit{#1}]}}
\else
\newcommand{\sji}[1]{}
\newcommand{\satish}[1]{}
\newcommand{\sli}[1]{}
\fi

\begin{document}
%
\title{Parallelizing Word2Vec in Shared and Distributed Memory}

\author{\IEEEauthorblockN{Shihao Ji, Nadathur Satish, Sheng Li, Pradeep Dubey}
\IEEEauthorblockA{Parallel Computing Lab, Intel Labs, USA\\
Emails: \{shihao.ji, nadathur.rajagopalan.satish, sheng.r.li, pradeep.dubey\}@intel.com}}

\maketitle

\begin{abstract}
Word2Vec is a widely used algorithm for extracting low-dimensional vector representations of words. It generated considerable excitement in the machine learning and natural language processing (NLP) communities recently due to its exceptional performance in many NLP applications such as named entity recognition, sentiment analysis, machine translation and question answering. State-of-the-art algorithms including those by Mikolov \emph{et al.} have been parallelized for multi-core CPU architectures but are based on vector-vector operations that are memory-bandwidth intensive and do not efficiently use computational resources. In this paper, we improve reuse of various data structures in the algorithm through the use of minibatching, hence allowing us to express the problem using matrix multiply operations. We also explore different techniques to distribute \wv computation across nodes in a compute cluster, and demonstrate good strong scalability up to 32 nodes. In combination, these techniques allow us to scale up the computation near linearly across cores and nodes, and process hundreds of millions of words per second, which is the fastest \wv implementation to the best of our knowledge. 
\end{abstract}


\IEEEpeerreviewmaketitle

\section{Introduction}
\label{sec:intro}

Natural language processing (NLP) aims to process text efficiently and enable understanding of human languages; it is one of the most critical tasks toward artificial intelligence \cite{ManningHinrich99}. One of the fundamental issues of NLP concerns
how machines can represent words of a language, upon which more
complex learning and inference tasks can be built efficiently. Instead
of the traditional bag of words (or one-hot) representation, distributed word
embedding represents each word as a dense vector in a low-dimensional embedding space 
such that semantically and syntactically similar words are close to each other.
This idea has been applied to a wide range of NLP tasks with considerable success \cite{CollobertWeston08,GloBorBen11,Turney13}.

Recently, Mikolov \emph{et al.} \cite{MikSutChe13} generated
considerable excitement in the machine learning and NLP communities by
introducing a neural network based model to learn distributed word
representations, which they call \wv. It was shown that \wv produces state-of-the-art performance on word similarity, word analogy tasks as well as many downstream NLP applications such as named entity recognition, machine translation and question answering~\cite{ChoMerGul14,WesChoBor15}.
The word similarity task is to retrieve words that are similar to a given word. On the other hand, word analogy requires answering queries of the form \emph{a:b;c:?}, where
\emph{a, b}, and \emph{c} are words from the vocabulary, and the answer
to the query must be semantically related to \emph{c} in the same way as
\emph{b} is related to \emph{a}. This is best illustrated with a
concrete example: Given the query \emph{king:queen;man:?} we expect the
model to output \emph{woman}.

The goal behind \wv is to find word representations that are useful for predicting the
surrounding words in a sentence. A common approach is to use
the Skip-gram model architecture with negative sampling~\cite{MikSutChe13}. This
method involves judging similarity between two words as the dot
product of their word representations, and the goal is to minimize the
distance of each word with its surrounding words while maximizing the
distances to randomly chosen set of words (a.k.a ``negative samples") that
are not expected to be close to the target.

The formulation of \wv uses Stochastic Gradient Descent (SGD) to solve this optimization problem. SGD solves optimization problems
iteratively; at each step, it picks a pair of words: an input word and another word either from its neighborhood or a random negative sample. It then
computes the gradients of the objective function with respect to the two
chosen words, and updates the word representations of the two words based on the gradient values. The algorithm then proceeds to the next iteration with a different word pair being chosen.

The formulation above has two main issues:

(1) SGD is inherently sequential: since there is a dependence between
the update from one iteration and the computation in the next
iteration (they may happen to touch the same word representations), each iteration must potentially wait for the
update from the previous
iteration to complete. This does not allow us to use the parallel resources of the hardware.

(2) Even if the above problem is solved, the computation performed in
each iteration is a single dot product of two word vectors. This is a level-1
BLAS \cite{BlaDemDon02} operation and is limited by memory bandwidth, thus not utilizing
the increasing computational power of modern multi-core and many-core
processors.

To solve (1), \wv uses Hogwild~\cite{NiuRecRe11}, a scheme where different
threads process different word pairs in parallel and ignore any
conflicts that may arise in the model update phases. In cache-coherent
architectures, however, Hogwild tends to have true and false sharing
of the model data structure between threads, and is heavily limited by
inter-thread communication.

In this work, we propose a simple yet efficient parallel algorithm to speed up the \wv computation in shared memory and distributed memory systems.
\begin{itemize}
\item We present a scheme based on minibatching and shared negative samples to convert the level-1 BLAS operations of \wv into the level-3 BLAS \cite{BlaDemDon02} matrix multiply operations, hence efficiently leveraging the vector units
and multiply-add instructions of modern architectures.
This is described in
Section~\ref{sec:algorithm}.

\item We parallelize this approach across batches of inputs, thereby
reducing the total number of model updates to the shared model and
hence limiting inter-thread communication. This allows our scheme to
scale better than Hogwild.

\item We perform experiments to scale out our technique
to a cluster of 32 compute nodes. Nodes across the cluster perform
synchronous model updates, and we follow the technique proposed
in ~\cite{CanZhaChe15} to reduce network traffic. 
We adjust the frequency
of propagating model updates across the network to achieve balance
between computation and communication. To maintain a good convergence rate in the presence of a limited number of updates (as number of compute nodes increases), we explore a simple learning rate adjustment trick without the computation and memory overheads of other learning rate scheduling techniques, such as AdaGrad \cite{DucHazSin11} and RMSProp \cite{Hinton12}.
\end{itemize}
In combination, these techniques allow us to scale the computation near linearly across cores and nodes, and process hundreds of millions of words per second, which is the fastest \wv implementation to the best of our knowledge.

The remainder of the paper is organized as follows. In Sec.~\ref{sec:model} we describe the basic \wv model and the original parallelization scheme proposed by Mikolov \emph{et al.}~\cite{MikSutChe13}. A new parallelization scheme is then presented in Sec.~\ref{sec:algorithm}, along with a distributed implementation cross nodes in a compute cluster. Example results on the One Billion Words Benchmark \cite{CheMikSch14} are presented in Sec.~\ref{sec:experiments}, with comparisons to the best known performance reported currently in the literature. Conclusions and future work are discussed in Sec.~\ref{sec:conclusion}.

\section{The Word2Vec Model}
\label{sec:model}

Word2vec represents each word $w$ in a vocabulary $V$ as a low-dimensional dense vector $\vb_w$ in an embedding space $\RR^D$, and attempts to learn the continuous word vectors $\vb_w$, $\forall w\in V$, from a training corpus such that the spatial distance between words then describes the similarity between words, e.g., the closer two words are in the embedding space, the more similar they are semantically and syntactically. These word representations are learned based on \emph{the distributional hypothesis}~\cite{MillerCharles91}, which assumes that words with similar context tend to have a similar meaning. Under this hypothesis, two distinct model architectures: Contextual Bag-Of-Words (CBOW) and Skip-Gram with Negative Sampling (SGNS) are proposed in \wv to predict a target word from surrounding context~\cite{MikCheCorDea13,MikSutChe13}. We focus here on the SGNS model since it produces state-of-the-art performance and is widely used in the NLP community. 

\begin{figure}[htb]
\centering
\includegraphics[width=3.2in]{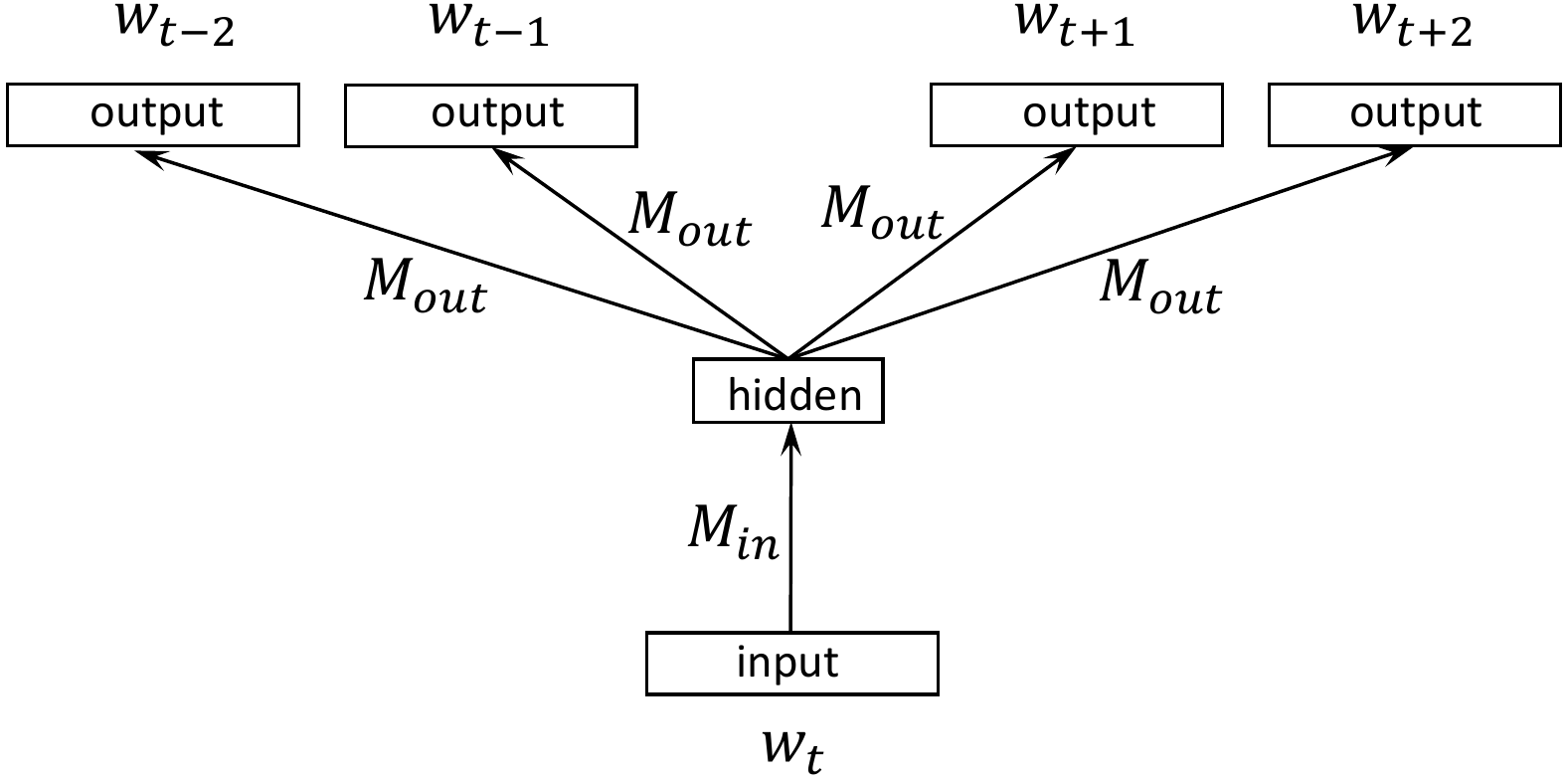} 
\caption{The Skip-gram model architecture based on a simplified one hidden layer neural network model.}
\label{fig:sgns}
\end{figure}
 
The training objective of the Skip-gram is to find word representations that are useful for predicting the surrounding words in a sentence from a large textual corpus. Given a sequence of training words $\{w_1,w_2,\cdots,w_T\}$, the objective of the Skip-gram model is to maximize the average log probability 
\begin{align}
  J(\Omega) = \frac{1}{T}\sum_{t=1}^T\sum_{-c\leq j\leq c, j\neq 0}\log p(w_{t+j}|w_t), 
  \label{eq:obj}
\end{align}
where $\Omega$ is the model parameters to be optimized (and will be defined soon), $c$ is the size of the training context (a sliding window around the center word $w_t$), and $p(w_{t+j}|w_t)$ is the probability of seeing word $w_{t+j}$ given the center word $w_t$. This probability function is formulated as a simplified one hidden layer neural network model, depicted in Fig.~\ref{fig:sgns}. The network has an input layer, a hidden layer without nonlinear transformation (also called projection layer), and a few softmax output layers, each of which corresponds to an output word within the context window. Typically, the network is fed as input $w_t\in\RR^{V}$, where $V$ denotes the vocabulary size, and it produces a hidden state $h \in\RR^{D}$, where $D$ is the size of the hidden layer or the dimension of the embedding space, which is in turn transformed to the output $w_{t+j}\in \RR^{V}$. Different layers are fully connected, with the weight matrix $M_{out}$ at output layers shared among all output words. Collecting all the weight matrices from this architecture, we denote the model parameter by $\Omega=\{M_{in}^{V\times D},M_{out}^{V\times D}\}$.

In the Skip-gram model above, the input $w_t$ is a sparse vector of a 1-of-$V$ (or one-hot) encoding with the element corresponding to the input word $w_t$ being 1 and the rest of components set to 0. Therefore, the basic Skip-gram formulation defines $p(w_{t+j}|w_t)$ as the softmax function:
\begin{align}
  p(w_O|w_I)=\frac{\exp(\inner{\vb_{in}^{w_I}}{\vb_{out}^{w_O}})}{\sum_{w=1}^V\exp(\inner{\vb_{in}^{w_I}}{\vb_{out}^{w}})}
  \label{eq:softmax}
\end{align}
where $\inner{\cdot}{\cdot}$ denote the inner product between two vectors, $\vb_{in}^w$ and $\vb_{out}^w$ are the ``input" and ``output" vector representations of $w$, corresponding to the respective rows of model parameter matrices $M_{in}$ and $M_{out}$. The computation of this formulation is prohibitively expensive since its cost is proportional to $V$, which is the size of the vocabulary and is often very large (e.g., around $10^6$). 

To improve performance of \wv, Mikolov \emph{et al.}~\cite{MikSutChe13} introduced negative sampling that approximates the log of softmax (\ref{eq:softmax}) as
\begin{align}
  \log p(w_O|w_I) &\approx \log\sigma(\inner{\vb_{in}^{w_I}}{\vb_{out}^{w_O}}) \nonumber\\
  &+ \sum_{k=1}^K\EE_{w_k\sim P_n(w)}[\log\sigma(-\inner{\vb_{in}^{w_I}}{\vb_{out}^{w_k}})], 
  \label{eq:sgns_obj}
\end{align}
where $\sigma(x)=\frac{1}{1+\exp(-x)}$ is the sigmoid (logistic) function, and the expectations are computed by drawing
random words from a sampling distribution $P_n(w)$, $\forall w\in V$. Typically the number of negative samples $K$ is much smaller than $V$ (e.g., $k\in[5, 20]$), and hence roughly a $V/K$ times of speed-up.

Even though negative sampling is an effective approximation technique, as the size of the corpus is typically at the order of billions of words and vocabulary size is at the order of millions (e.g., $T=10^9$, and $V=10^6$), training \wv model often takes tens of hours or even days for some Internet scale applications.

\section{Word2vec Algorithm and improvements}
\label{sec:algorithm}
In order to solve the optimization problem described in the previous
section, Stochastic Gradient Descent (SGD) is commonly used.
SGD is an iterative algorithm; at each
iteration, a single $(w_I, w_O)$ pair is picked, where $w_I$ is an input
context word and $w_O$ is a target word or a negative sample. The gradient of
the objective function is then calculated w.r.t. the word
vectors for $w_I$ and $w_O$; and a small change/update is made to these
vectors. One of the problems of SGD is that it is inherently challenging to parallelize, i.e., SGD only
updates the word vectors of a pair of words at a time, and parallel model
updates on multiple threads can result in conflicts if the threads try
to update the vectors of the same word.

\begin{figure*}[htb]
\centering
\includegraphics[width=5.5in]{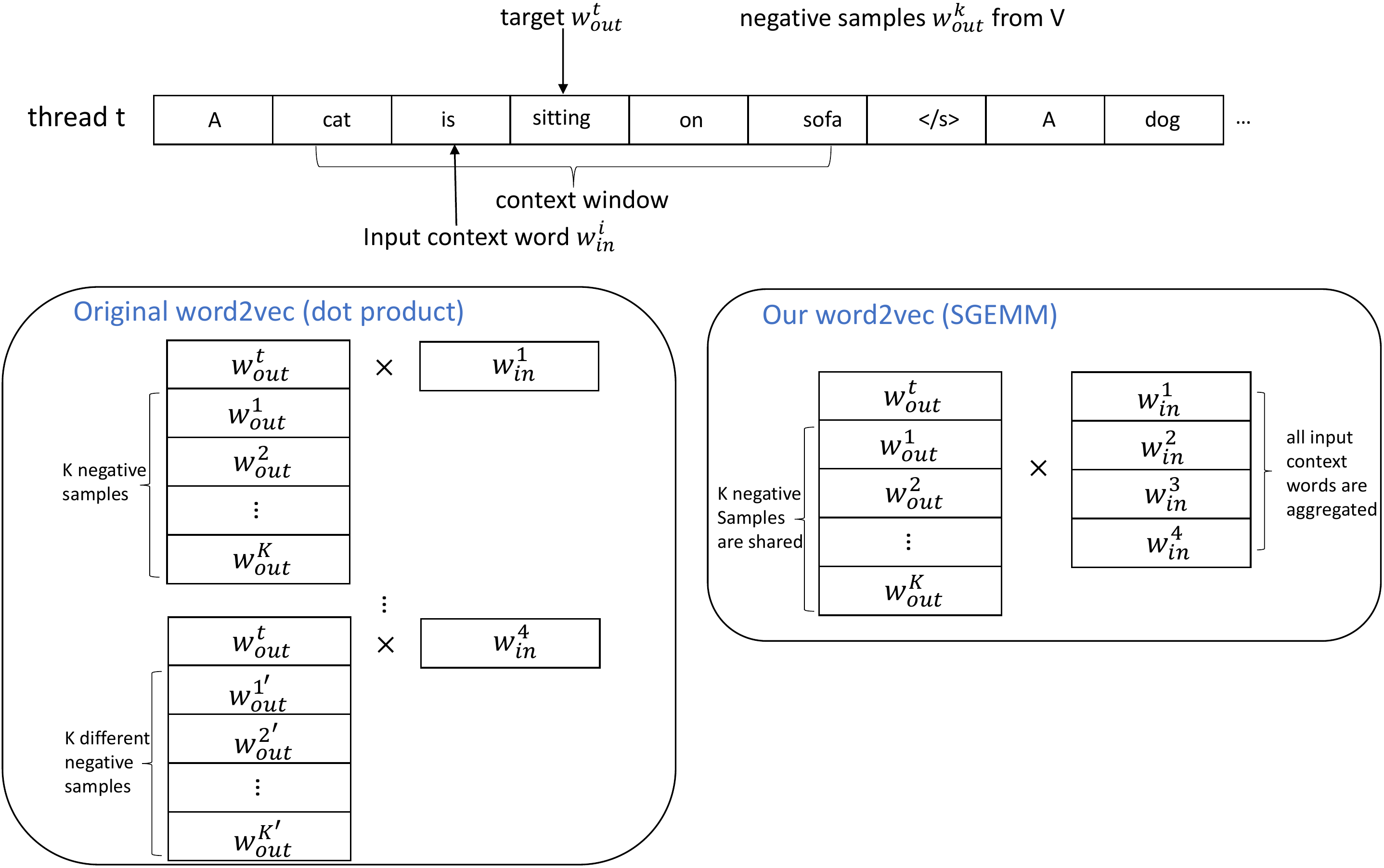} 
\caption{The parallelization schemes of the original word2vec (left) and our optimization (right).}
\label{fig:schemes}
\end{figure*}

The original implementation of \wv by Mikolov \emph{et
al.}~\footnote{\url{https://code.google.com/archive/p/word2vec/}} uses
Hogwild~\cite{NiuRecRe11} to parallelize SGD.
Hogwild is a parallel SGD algorithm that seeks to ignore
conflicts between model updates on different threads and allows updates to proceed even in the presence of conflicts.
The psuedocode of Hogwild SGD update is shown in
Algorithm~\ref{alg:hogwild}. The algorithm takes in a matrix $M_{in}^{V\times D}$ that
contains the word representations for each input word, and a matrix $M_{out}^{V\times D}$
for the word representations of each output word.
Each word is represented as an array of $D$ floating point numbers,
corresponding to one row of the two matrices.
These matrices are updated during the computation. We also take in a
specific target word, and a set of $N$ input context words around the target as
depicted in Fig.~\ref{fig:schemes}.
The algorithm iterates over the $N$ input words in Lines 2-3.
The psuedocode only shows a single thread; in Hogwild, the loop
in Line 2 is parallelized over threads without any additional change
in the code. In the loop at Line 6, we pick either the positive
example (the target word in Line 8) or a negative example at random (Line
10). Lines 13-15 compute the gradient of the objective function with
respect to the choice of input word and positive/negative example. Lines
17--20 perform the update to the entries $M_{out}[\text{pos/neg example}]$
and $M_{in}[\text{input context}]$.

\begin{table}
\captionof{algorithm}{Hogwild SGD implementation of \wv in one thread.}
\label{alg:hogwild}
\begin{minipage}{0.01\linewidth}\hspace{0.01pt}\end{minipage}
\begin{minipage}{0.95\linewidth}
\begin{tabular}{l}
\begin{lstlisting}
@Given model parameter $\Omega=\{M_{in},M_{out}\}$, learning rate $\alpha$, 1 target word $w_{out}^t$, and N input words  \{$w_{in}^0$, $w_{in}^1$, $\cdots$, $w_{in}^{N-1}$\}@
for (i = 0; i < N; i++) {
  input_word = @$w_{in}^i$@;
  for (j = 0; j < D; j++) temp[j] = 0;
  // negative sampling
  for (k = 0; k < negative + 1; k++) {
    if (k = 0) {
      target_word = @$w_{out}^t$@; @label@ = 1;
    } else {
      target_word = sample one word from V; @label@ = 0;
    }
    inn = 0;
    for (j = 0; j < D; j++) inn += @$M_{in}$@[input_word][j] * @$M_{out}$@[target_word][j];
    err = @label@ - @$\sigma$@(inn);
    for (j = 0; j < D; j++) temp[j] += err * @$M_{out}$@[target_word][j];
    // update output matrix
    for (j = 0; j < D; j++) @$M_{out}$@[target_word][j] += @$\alpha$@ * err * @$M_{in}$@[input_word][j];
  }
  // update input matrix
  for (j = 0; j < D; j++) @$M_{in}$@[input_word][j] += @$\alpha$@ * temp[j] ;
}
\end{lstlisting}\\
\end{tabular}
\end{minipage}
\end{table}

Algorithm~\ref{alg:hogwild} reads and updates
entries corresponding to the input context and positive/negative words at
each iteration of the loop at Line 6. This means that there is a
potential dependence between successive iterations. Hogwild ignores
such dependencies and proceeds with updates regardless of conflicts.
In theory, this can reduce the rate of convergence of the algorithm as
compared to a sequential run. However, the Hogwild approach has been
shown to work well in case the updates across
threads are unlikely to be to the same word; and indeed for large
vocabulary sizes, conflicts are relatively rare and convergence is not
typically affected.

\subsection{Advantages and Drawbacks of Algorithm~\ref{alg:hogwild}}
Algorithm~\ref{alg:hogwild} has a few main advantages: threads do not
need to synchronize between updates and can hence proceed
independently with minimal instruction overheads. Further, the
computation of the gradient is based off the current state of the
model visible to the thread at that time. Since all threads update the
same shared model, the values read are only as stale as the
communication latency between threads, and in practice this does not
cause much convergence problems for \wv.

However, the algorithm suffers from two main drawbacks that
significantly affect runtimes. First, since
multiple threads can update the same cache line containing a specific
model entry, there can be significant ping-ponging of cache lines
across cores. This leads to high access latencies and significant drop
in scalability. Second and perhaps even more importantly, there is
a significant amount of locality in the model updates that is not
exploited in the Hogwild algorithm. As an example, we can easily see
that the same target word is used in the model updates for several
input words. By performing a single update at a time, this locality
information is lost, and the
algorithm performs a series of dot-products that are level-1 BLAS operations and limited
by memory bandwidth. It is indeed, as we show next, possible to batch
these operations into a level-3 BLAS call which can more efficiently utilize
the compute capabilities and the instruction sets of modern multi-core
and many-core architectures.

\subsection{New Parallelization Scheme in Shared Memory}

We first discuss how we can exploit the available locality in
Algorithm~\ref{alg:hogwild}. This can be done even on a single compute
thread. We then describe the impact of this step on parallelization
and inter-thread communication.

We exploit locality in two steps. As a motivation, consider
Fig.~\ref{fig:schemes}. The figure to the left shows the
parallelization scheme of the original \wv. Note that we compute
dot products of the word vectors for a given input word $w_{in}^{i}$
with both the target word $w_{out}^{t}$ as well as a set of $K$ negative samples
$\{w_{out}^{1}, \cdots, w_{out}^{K}\}$. Rather than doing these one at a
time, it is rather simple to batch these dot products into a matrix
vector multiply, a level-2 BLAS operation, as shown in the left side of Fig.~\ref{fig:schemes}.
However, this alone does not buy significant performance improvement.
Indeed, most likely the shared input word vector may come from cache.

In order to convert this to a level-3 BLAS operation, we also need to batch
the input context words. Doing this is non-trivial since the negative
samples for each input word could be different in the original
\wv implementation. We hence propose ``negative sample sharing'' as a strategy,
where we share negative samples across a small batch of input words.
Doing so allows us to convert the original dot-product based
multiply into a matrix-matrix multiply call (GEMM) as shown on the right side of
Fig.~\ref{fig:schemes}. At the end of the GEMM, the model
updates for all the word vectors of all input words and target/sample
words that are computed need to be written back. Performing
matrix-matrix
multiplies (GEMMs) rather than dot-products allows us to leverage
all the compute capabilities of modern architectures including
instruction set features such as multiply-add instructions in the
Intel 
AVX2 instruction set.
It also allows us to leverage heavily optimized linear algebra
libraries. Note that typical matrix dimensions are not very large. For
instance, the number of negative samples is only 5--20, and the batch
size for the input batches are limited to about 10--20 for convergence
reasons. Nevertheless, we find that we get considerable speedups even
with this level of reuse over the original \wv.

\subsection{Consequence of the new parallelization scheme}
While the original \wv performs model updates (and potentially the inter-thread
communication that comes with it) after each dot product, our new parallelization
scheme above performs a number of dot products as a GEMM call
(corresponding to multiple input words and multiple samples) before
performing model updates. We follow a simple ``Hogwild"-style
philosophy for multi-threading across the GEMM calls - we
allow for threads to potentially conflict when updating the models at
the end of the GEMM operation.

It is important to note that the locality optimization has a secondary
but important benefit - we cut down on the total number of updates to the model.
This happens since the GEMM operation performs a reduction
(in registers/local cache) to an update to a single entry in the
output matrix; while in the original \wv scheme such updates to the same
entry (same input word representation, for instance) happen at
distinct periods of time with potential ping-pong traffic happening
in between. As we will see in Sec.~\ref{sec:experiments} when we present results, this leads to
a much better scaling of our approach than the original \wv.


However, we need to pay careful emphasis to the rate of convergence while doing these
transformations. In contrast to the original \wv that does small partial model
updates frequently, our new GEMM based scheme batches many model updates
together and performs less frequent updates. This can result in
different multi-threading behavior. Specifically, it is possible that
threads read a more up-to-date model in the original \wv as opposed to the GEMM
based scheme. The extent to which this occurs is, of course, dependent
on the batch size we use for the inputs. In our experiments, with a
batch size of about 10-20, we have not found any significant impact on
convergence.
One reason for this is that many of the intermediate model updates in
the original \wv are to parts of the model that will be updated again in the
very near future -- for example, updates to the same input word due to multiple same input words occur close by in time. Even if relatively
updated models are seen in the original \wv, they are still not the final
result and could be partially updated due to model update conflicts from multiple threads.

\subsection{Comparison to BIDMach}
\label{sec:bidmach}
The \wv implementation in BIDMach~\cite{CanZhaChe15} also uses
the previously described idea of shared negative samples. However,
the computation in BIDMach is organized in a different way. First,
BIDMach separates the handling of the positive examples and negative
samples into two steps. For handling positive examples, BIDMach 
iterates over each word and performs dot products of word vectors
considering that word as the target and surrounding words as input context.
We can think of these operations as a sequence of matrix vector
products, each time with a single target and 
corresponding input context words. There is some reuse of context words
across matrix vector calls due to the overlap in context between
successive target words. However, since computation is not batched
into higher level BLAS calls, BIDMach cannot fully exploit this reuse 
through standard techniques such as register and
cache blocking -- register and cache state may not be maintained across loop
iterations. In a similar way, BIDMach also processes negative samples
as a sequence of dot products, and suffers from similar limitations. 
In contrast, we directly exploit reuse of input context words across the
positive and negative samples using a GEMM call. The underlying
optimized libraries can then exploit reuse across all levels of the
register and cache hierarchy. We demonstrate the performance impact of
both designs when we present results in Sec.~\ref{sec:experiments}.



\subsection{Distributed Memory Parallelization}
Scalability on multi-node distributed system is as important as, if not more important than, that on single node system. This is because typical large scale machine learning applications are compute intensive and require days, weeks even months of training time. In the case of \wv, even with the techniques we proposed above, it still takes tens of hours or even days to train on some of the largest data sets in the industry, such as the 100 billion word news articles from Google. Thus, scaling out \wv on multi-node distributed system is critical in practice.

To scale out \wv, we explore different techniques
to distribute its computation across nodes in a compute cluster.
Since the individual matrix multiplies are not very large, there is
not too much performance that can be gained from distributing these
across multiple nodes (a.k.a. model parallelism).
Therefore, data parallelism is considered for distributed
implementation. In data parallelism with $N$ computing nodes, the
training corpus is equally partitioned into $N$ shards and the model parameters
$\Omega=\{M_{in},M_{out}\}$ are replicated on each computing node; each
node then independently processes the data partition it owns and updates its local model, and periodically synchronizes the local model with
all the other $N-1$ nodes.

There are two common issues to be addressed in data parallelism: (1)
efficient model synchronization over the communication network, and
(2) improving the statistical efficiency of large mini-batch SGD. The
first issue arises because typical network bandwidths are an order of
magnitude lower than CPU memory bandwidths. For example, in
commonality cloud computing infrastructures such as AWS the network
bandwidths are around 1GB/sec; even in HPC system with FDR infiniband,
the network bandwidths are still of the order of 10GB/sec. As the
typical size of the model $\Omega$ is about 2.5GB in our experiments,
full model synchronization over 4 computing nodes connected via FDR
Infiniband takes about 0.5 seconds, which is too slow to keep up with
local model updates. In the case of \wv, however, not all word vectors are updated at the same frequency as those are proportional to the word unigram frequencies, e.g., the vectors in the model associated with popular words are updated more frequently than those of rare words. We therefore strive to match model update frequency to word frequency, and a sub-model (instead of full-model) synchronization scheme, similar to the one exploited in BIDMach~\cite{CanZhaChe15}, is used.


The second issue arises because as the number of nodes $N$ increases, conceptually a $N$ times larger mini-batch is used in SGD update, which affects the statistical efficiency and slows down the rate of convergence. Fortunately, this issue has been studied recently and various techniques are proposed to mitigate the loss of convergence rate. We follow the $m$-weighted sample scheme studied in Splash~\cite{ZhangJordan15} and increase the starting learning rate as the number of nodes increases while exploring different learning rate scheduling techniques, such as AdaGrad \cite{DucHazSin11} and RMSProp \cite{Hinton12}, to improve convergence rate. From our experiments, we found that while AdaGrad and RMSProp are effective techniques to speed up convergence, they incur large memory consumption since they dedicate a learning rate to each model parameter and need a separate matrices of the same size as $\Omega$ to store the per-parameter learning rates. In addition, accessing large memory arrays makes the algorithm memory-bandwidth intensive and slows down the throughput considerably. Instead, we found that a simple learning rate update schedule based on a single learning rate is quite satisfactory, and empirically we note that we just need to reduce the learning rate more aggressively as number of nodes increases. We demonstrate the effectiveness of these techniques in our experiments next.

\section{Experiments}
\label{sec:experiments}

We optimize \wv with the techniques discussed above both in single node shared memory system and in multi-node distributed system. We report the \emph{system-performance} measured as throughput, i.e., million words/sec, and the \emph{predictive-performance} measured as accuracy on standard word similarity and word analogy test sets. The performances of our optimization are compared with the original \wv on CPUs, and with the state-of-the-art results reported in literature on Nvidia GPUs. Our code will be made available for general usage.



\subsection{Experimental Setup}

\textbf{Hardware}: The majority of our experiments are performed on
two Intel architectures for shared memory and distributed memory computation: (1) dual-socket Intel Xeon 
{
E5-2697 v4 Broadwell CPUs, and (2) the latest Intel Xeon Phi Knights Landing (KNL) processors. The Broadwell processor has 36 cores (72 threads including Simultaneous Multi-Threading/SMT) running at 2.3 GHz, and the KNL processor has 68 cores and each core has 4 hardware threads (or 272 threads in total) running at 1.4 GHz. Each machine has 128 GB RAM and runs Red Hat Enterprise Linux Server release 6.5. In the distributed setting, all the Broadwell nodes are connected through FDR infiniband, and all the KNL nodes are connected through Intel Omni-Path (OPA) Fabric.

\textbf{Software}: We use custom end-to-end code written in C++ with
OpenMP, and compiled with the Intel 
C++ Compiler version 16.0.2. 
We use Intel 
MKL version 11.3.2 and Intel
MPI library version 5.1.3 for SGEMM calls and multi-node massage passing.

\textbf{Training corpora}: We train our \wv models on three different corpora: (1) a small (text8) dataset\footnote{\url{http://mattmahoney.net/dc/text8.zip}} of 17 million words from wikipedia that is widely used for word embedding demos, (2) the recently released One Billion Words benchmark~\cite{CheMikSch14}, and (3) a large collection of 7.2 billion words that we gathered from a variety of data sources: the 2015 Wikipedia dump with 1.6 billion words, the WMT14 News Crawl\footnote{\url{http://www.statmt.org/wmt14/translation-task.html}} with 1.7 billion words, the aforementioned one billion word benchmark, and UMBC webbase corpus\footnote{\url{http://ebiquity.umbc.edu/resource/html/id/351}} with around 3 billion words. Different corpora are used in order to verify the generalization performance of our algorithm under different training data statistics. The one billion word benchmark~\cite{CheMikSch14} is our main dataset for throughput and predictive accuracy study since this is the benchmark on which the best known GPU performances were reported.

\textbf{Test sets}: The quality of trained models are evaluated on
\emph{word similarity} and \emph{word analogy} tasks. For word
similarity, we use WS-353~\cite{FinGabMat02} which is one of the most
popular test datasets used for this purpose. It contains word pairs together with human-assigned
similarity judgments. The word representations are evaluated by ranking the pairs according to
their cosine similarities, and measuring the Spearman’s rank correlation coefficient with the human
judgments. For word analogy, we use the Google analogy dataset~\cite{MikCheCorDea13}, which contains 19544 word analogy questions, partitioned into 8869 semantic and 10675 syntactic questions. The semantic questions contain five types of semantic analogies, such as capital cities (Paris:France;Tokyo:?), currency (USA:dollar;India:?) or people (king:queen;man:?). The syntactic questions contain nine types of analogies, such as plural nouns, opposite, or comparative, for example good:better;smart:?. A question is correctly answered only if the algorithm selects the word that is exactly the same as the correct word in the question.

\textbf{Code}: We compare the performances of three different implementations of \wv: (1) the original implementation from Google that is based on Hogwild SGD on shared memory systems (\url{https://code.google.com/archive/p/word2vec/}), (2) BIDMach (\url{https://github.com/BIDData/BIDMach}) which achieves the best known performance of \wv on Nvida GPUs, and (3) our optimized implementation on Intel architectures. 

\textbf{Word2vec parameters}: In the experiments on the one billion word benchmark, we follow the parameter settings of BIDMatch (dim=300, negative samples=5, window=5, sample=1e-4, vocabulary of 1,115,011 words). In this case, the size of the model $\Omega=\{M_{in}, M_{out}\}$ is about 2.5GB. Similar parameter settings are used for the small text8 dataset and the 7.2 billion word collection discussed above.

\subsection{Single Node Shared Memory Systems}
To achieve high performance on modern multi-socket multi-core shared memory systems, parallel algorithms need to have strong scalability across cores and sockets. Scaling across cores is challenging for \wv because more threads creates more inter-core traffic due to cache line conflicts (including false sharing), which prevents it from achieving good scalability. Scaling across sockets is even more challenging since the same traffic caused by cache line conflicts and false sharing needs to travel across sockets. The high inter-socket communication overhead imposes a major hurdle to achieve good scalability across sockets.

\textbf{System-Performance (Throughput)}: Fig.~\ref{fig:core_scaling} shows the \emph{system-performance} measured as million words/sec of our algorithm and the original \wv, scaling across all cores/threads and sockets of a 36-core dual-socket Intel Broadwell CPU. We use the one billion word benchmark~\cite{CheMikSch14} in the experiment. When using only one thread, our optimization achieves 2.6X speedup over the original \wv. The superior performance of our optimization is due to the new parallelization scheme which is more hardware-friendly after converting level-1 BLAS dot-products to level-3 BLAS matrix multiplies as described in Sec.~\ref{sec:algorithm}.

\begin{figure}[htb]
\centering
\includegraphics[width=3.4in]{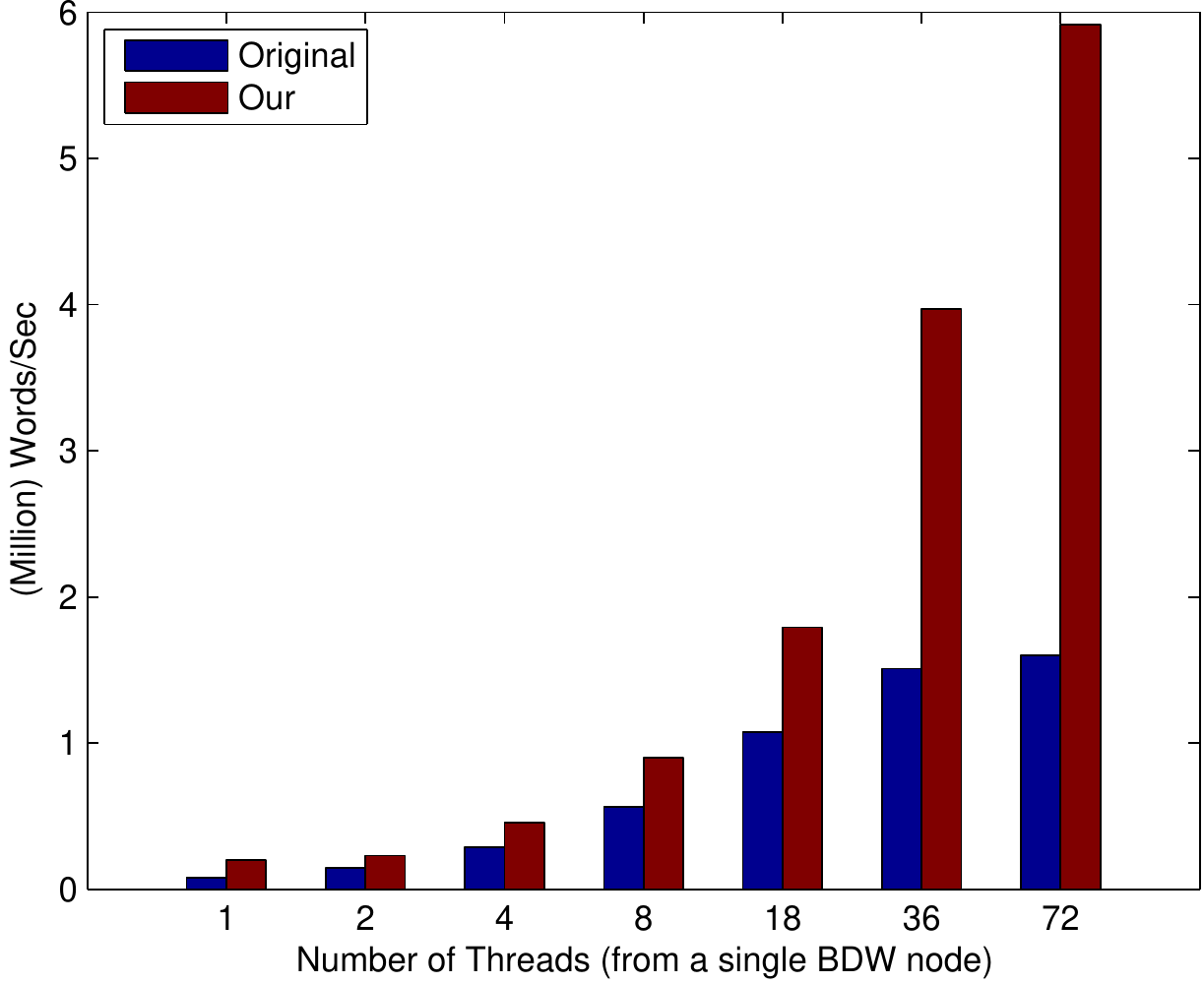} 
\caption{Scalabilities of the original \wv and our optimization on all threads of an Intel Broadwell CPU; evaluated  on the one billion word benchmark~\cite{CheMikSch14}.}
\label{fig:core_scaling}
\end{figure}


When scaling to multiple threads, our algorithm achieves linear speedup as shown in Fig.~\ref{fig:core_scaling}. This linear scalability is near perfect within a single socket (when number of threads $\leq 36$), and the scalability becomes sub-linear when two sockets are involved (when number of threads = 72) in which case cross-socket memory access penalizes the potential linear scaling. In contract, the original \wv scales linearly only until 8 threads and slows down significantly after that. In the end, the original \wv delivers about 1.6 million words/sec, while our code delivers 5.8 million words/sec or a 3.6X speedup over the original \wv. The superior performance highlights the effectiveness of our optimization, as compared to the original \wv, in reducing unnecessary inter-thread communications and utilizing computation resource of modern multi-core architecture.

\textbf{Predictive-Performance (Accuracy)}: Delivering higher throughput is only meaningful when the trained model reaches similar or better predictive accuracy. We therefore evaluate the models trained from the original \wv and our implementation, and report their predictive performances on the word similarity and word analysis tasks in Table~\ref{tab:accuracy_shared}. In order to verify the generalization performance of our techniques, we run the respective codes on three different training corpora as described above.

\begin{table}[htb]
\caption {Predictive performances of the models trained from the original \wv and our optimization on three different training corpora. All the experiments are performed on an Intel Broadwell CPU.}
\label{tab:accuracy_shared}
\begin{center}
\begin{tabular}{|l|r|c|c|c|c|}
\hline         & Vocabulary  & \multicolumn{2}{c|}{Word Similarity} & \multicolumn{2}{c|}{Word Analogy} \\
\cline{3-6}  \multicolumn{1}{|c|}{Corpus} & \multicolumn{1}{c|}{Size} & Original & Our        & Original & Our \\
\hline          17M-word (text8)        & 71,291 & 63.4 & 66.5 & 17.2 & 18.1 \\
\hline          1B-word benchmark    & 1,115,011 & 64.0  & 64.1 & 32.4 & 32.1 \\
\hline          7.2B-word collection & 1,115,011 & 70.0 & 69.8 & 73.5   & 74.0 \\
\hline
\end{tabular}
\end{center}
\end{table}

\begin{table}[htb]
\caption {Predictive performances of trained models on one billion word benchmark with vocabularies of different sizes.}
\label{tab:accuracy_vocab}
\begin{center}
\begin{tabular}{|r|c|c|c|c|}
\hline         Vocabulary  & \multicolumn{2}{c|}{Word Similarity} & \multicolumn{2}{c|}{Word Analogy} \\
\cline{2-5}   \multicolumn{1}{|c|}{Size} & Original & Our        & Original & Our \\
\hline          1,115,011 & 64.2 & 64.1 & 32.4 & 32.1 \\
\hline          500,000   & 63.0 & 62.4 & 32.2 & 33.0 \\
\hline          250,000   & 63.1 & 61.8 & 32.2 & 33.0 \\
\hline          100,000   & 55.6 & 55.8 & 32.2 & 31.9 \\
\hline           50,000   & 49.7 & 49.7 & 30.1 & 29.9 \\
\hline
\end{tabular}
\end{center}
\end{table}

As can be seen from Table~\ref{tab:accuracy_shared}, our code achieves very similar (most of time higher) predictive accuracy, compared to the original \wv, cross three different corpora. It demonstrates that our optimization generalizes very well to different corpora that have a variety of sizes under different vocabulary settings.


To examine the robustness of our \wv further, we study its predictive accuracy under varying data statistics. We again run the original \wv and our optimization on the one billion word benchmark but with vocabularies of different sizes. For the vocabulary of size $N$, we keep the top $N$ most popular words occurred in the corpus in the vocabulary. These popular words have the most of occurrences in the training corpus, and therefore their updates (and also the conflicts in the ``Hogwild"-style SGD) are more frequent than those on rare words. It can been seen from Table~\ref{tab:accuracy_vocab} that both the original \wv and our optimization achieve very similar accuracies for all vocabulary sizes, including the most challenging one with a small vocabulary of 50K words.

Overall, these experiments demonstrate that the parallization scheme and the optimization techniques we proposed in Sec.~\ref{sec:algorithm} delivers 3X-4X speedup over the original \wv without loss of predictive accuracy.

\textbf{Comparison to state-of-the-arts}: After demonstrating the superior performances of our optimization, we now perform detailed comparison to the state-of-the-arts, including the original \wv from Google and BIDMach. Since all the implementations achieve similar accuracy, we focus on the throughput in the comparison. Improving throughput (while maintaining accuracy) is always important since it democratizes the large \wv models by lowering the training costs. Thus, extensive studies have been focused on improving throughput. The best known performance reported currently in the literature is from BIDMach on the one billion word benchmark using Nvidia GPUs~\cite{CanZhaChe15}. We therefore run our experiments on the same benchmark using the same parameter setting as that of BIDMach. Moreover, to evaluate the generalization of our techniques, we also run our experiments on three different Intel architectures including the latest Intel Xeon Phi Knight Landing processor.

\begin{table}[htb]   
  \caption {Performance comparison of the state-of-the-art implementations of \wv on different architectures, including dual-socket 28-core Intel Haswell E5-2680 v3, dual-socket 36-core Intel Broadwell E5-2697 v4, single-socket 68-core Intel Knights Landing, Nvidia K40 GPU, and Nvidia GeForce Titan-X GPU. Results on CPU-platforms are obtained from our experiments, while results on the GPU systems are obtained from published literature~\cite{CanZhaChe15}. Results are evaluated on the one billion word benchmark~\cite{CheMikSch14}.} 
  \label{tab:throughputs_shared}
  \begin{center}
	\begin{threeparttable}
    \begin{tabular}{|l|l|c|}
      \hline Processor      & Code  & Words/Sec \\
      \hhline{|=|=|=|} Intel HSW (Xeon E5-2680 v3) & Original & 1.5M \\
      \hline Intel HSW (Xeon E5-2680 v3) & BIDMach & 2.4M \\
      \hline Intel HSW (Xeon E5-2680 v3) & Our & \textbf{4.2M} \\
      \hhline{|=|=|=|} Intel BDW (Xeon E5-2697 v4) & Original & 1.6M \\
      \hline Intel BDW (Xeon E5-2697 v4) & BIDMach & 2.5M \\
      \hline Nvdia K40 & BIDMach & \hspace{0.05in}4.2M$^1$ \\
      \hline Intel BDW (Xeon E5-2697 v4) & Our & \textbf{5.8M} \\
      \hhline{|=|=|=|} Nvdia GeForce Titan-X & BIDMach & \hspace{0.05in}8.5M$^1$ \\
      \hline Intel KNL (Xeon Phi) & Our & \textbf{8.9M} \\
      \hline
    \end{tabular}
    \begin{tablenotes}
  		\scriptsize\item $^1$Data from \cite{CanZhaChe15}.
	\end{tablenotes}
	\end{threeparttable}
  \end{center}
\end{table}

Table~\ref{tab:throughputs_shared} shows the detailed comparisons. On Intel Haswell and Broadwell architectures, BIDMach and our optimization outperform the original \wv: typically BIDMach delivers 1.6X speedup over the original \wv while our optimization delivers 2.8X-3.6X speedup. In addition, our performance on Intel Broadwell (5.8 million words/sec) outperforms BIDMach's performance on Nvidia K40 (4.2 million words/sec). The best known performance on shared memory system was reported by BIDMach \cite{CanZhaChe15} on Nvidia GeForce Titan-X (8.5 million words/sec) which is 1.5X faster than our performance on Intel Broadwell. However, in terms of compute efficiency, BIDMach on Nvidia Titan-X is much lower than our code on Intel Broadwell since the former has 3X peak flops of the latter, indicating that BIDMach's efficiency on Nvidia Titan-X is only half of ours on Intel Broadwell. This is likely due to the parallelization scheme of BIDMach, which cannot efficiently use all computational resources, as we discussed in Sec.~\ref{sec:bidmach}. Finally, our optimization on Intel KNL processor delivers 8.9 million words/sec, a new record on the one billion word benchmark achieved on a single node shared memory system.



\subsection{Distributed Multi-node Systems}

Next we demonstrate the scalability and predictive performance of our distributed \wv on multi-node distributed systems. The experiments with our distributed \wv are performed on two CPU clusters: (1) Intel Broadwell nodes connected via FDR Infiniband, and (2) Intel KNL nodes connected via Intel OPA Fabric. Fig.~\ref{fig:scaling} shows the scalability of our distributed \wv on both distributed systems as number of nodes increases, while Table~\ref{tab:accuracy_dist} reports the corresponding predictive performances on the word similarity and word analogy benchmarks. For the purpose of comparison, we also include in Fig.~\ref{fig:scaling} BIDMach's performances on $N=1,4$ NVidia Titan-X GPUs provided by \cite{CanZhaChe15}, which reports the state-of-the-art performance achieved on multi-GPU systems. Again, good scalability is only meaningful when similar or better accuracy is achieved. We therefore provide the predictive performance of the original \wv as the baseline in Table~\ref{tab:accuracy_dist}. 



\begin{figure}[htb]
\centering
\includegraphics[width=3.4in]{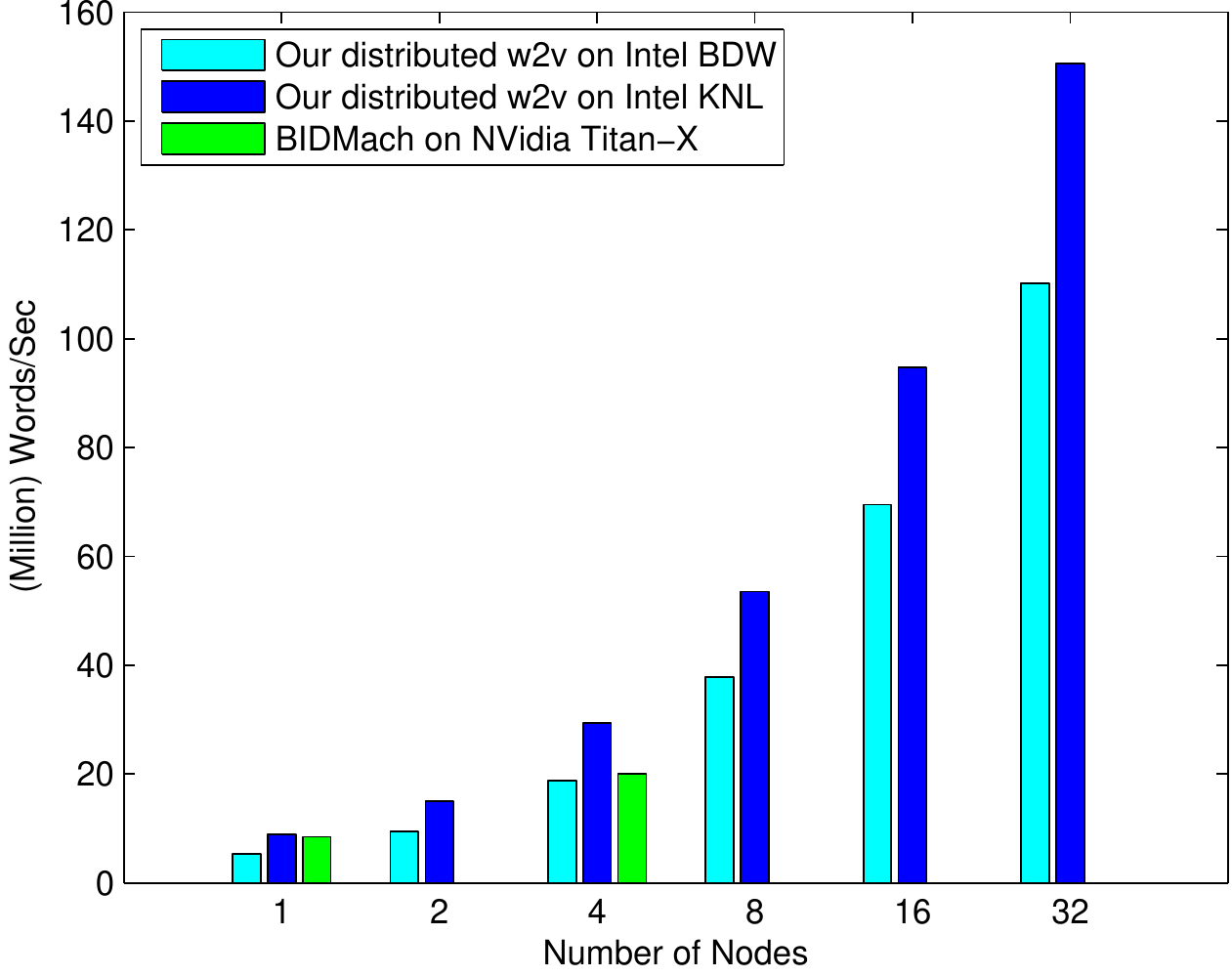} 
\caption{Scalabilities of our distributed \wv on multiple Intel Broadwell and Knight Landing nodes, and BIDMach on $N=1,4$ NVidia Titan-X nodes as reported in \cite{CanZhaChe15}.}.
\label{fig:scaling}
\end{figure}

\begin{table}[htb]
\caption {Predictive performances of our distributed \wv trained on the one billion word benchmark~\cite{CheMikSch14} evaluated on the word similarity and word analogy tasks. The performance of the original \wv is provided as baseline.}
\label{tab:accuracy_dist}
\begin{center}
\begin{tabular}{|l|c|c|c|c|}
\hline         \#Nodes  & \multicolumn{2}{c|}{Word Similarity} & \multicolumn{2}{c|}{Word Analogy} \\
\hline          Original ($N=1$) & \multicolumn{2}{c|}{64.0} & \multicolumn{2}{c|}{32.4} \\
\hline          Distributed w2v & BDW & KNL & BDW & KNL \\
\hline          \multicolumn{1}{|c|}{$N=1$}    & 64.1 & 63.7 & 32.1 & 32.1 \\
\hline          \multicolumn{1}{|c|}{$N=2$}    & 64.1 & 65.2 & 32.3 & 32.5 \\
\hline          \multicolumn{1}{|c|}{$N=4$}    & 63.0 & 63.4 & 32.0 & 32.2 \\
\hline          \multicolumn{1}{|c|}{$N=8$}    & 63.8 & 64.9 & 32.1 & 31.3 \\
\hline          \multicolumn{1}{|c|}{$N=16$}   & 62.8 & 62.3 & 31.6 & 31.0 \\
\hline          \multicolumn{1}{|c|}{$N=32$}   & 63.2 & 61.2 & 31.1 & 30.1 \\
\hline
\end{tabular}
\end{center}
\end{table}

As can been seen from Fig.~\ref{fig:scaling} and Table~\ref{tab:accuracy_dist}, our distributed \wv achieves near linear scaling until 16 Broadwell nodes or 8 KNL nodes while maintaining a comparable accuracy to that of the original \wv. As the number of nodes increases, to achieve the linear scaling while maintaining a comparable accuracy, we need to increase the learning rate and the model synchronization frequency slightly to mitigate the loss of convergence rate. When number of Broadwell nodes increases to 32 (or KNL for 16), we need to further increase model synchronization frequency to maintain a good predictive accuracy. However, the increment of model synchronization frequency takes a toll on the scalability, and leads to a sub-linear scaling at 32 Broadwell nodes or 16 KNL nodes. Despite of this, our distributed \wv delivers over 100 million words/sec with a small 1\% accuracy loss. To the best of our knowledge, this is the best performance reported so far on this benchmark. As a comparison, BIDMach only delivers 2.4X speedup on 4 GPU cards vs. 1 GPU card or a 60\% efficiency.

\begin{table}[htb]   
  \caption {Performance comparison of state-of-the-art distributed \wv trained on the one billion word benchmark~\cite{CheMikSch14} with multi-node CPU and GPU systems} 
  \label{tab:throughputs_dist}
  \begin{center}
	\begin{threeparttable}
    \begin{tabular}{|l|l|l|c|}
      \hline Systems   & Node Count   & Code  & Words/Sec \\
      \hhline{|=|=|=|=|} Nvidia Titan-X GPU & 4 nodes & BIDMach & \hspace{0.05in}20M$^1$ \\
      \hline Intel Broadwell CPU & 4 nodes & Our & 20M \\
      \hline Intel Knights Landing & 4 nodes & Our & 29.4M \\
      \hline Intel Broadwell CPU & 32 nodes & Our & 110M \\
      \hline Intel Knights Landing & 16 nodes & Our & 94.7M \\
      \hline
    \end{tabular}
    \begin{tablenotes}
  		\scriptsize\item $^1$Data from \cite{CanZhaChe15}.
	\end{tablenotes}
	\end{threeparttable}
  \end{center}
\end{table}


Last, we collect the best known performance of distributed
\wv from the literature \cite{CanZhaChe15}, and compare it with our performance on Intel Broadwell and KNL nodes and report them in Table~\ref{tab:throughputs_dist}. We only consider the meaningful throughputs that maintain a comparable accuracy. Therefore, only the performances of 32 Broadwell nodes and 16 KNL nodes are included. As can be seen, our 4 Broadwell nodes matches BIDMach's performance
on 4 Nvidia Titan-X cards, and we deliver about 110
million words/sec on a cluster of 32 Intel Broadwell nodes,
the best performance reported so far on this benchmark. With 16 Intel KNL nodes, we deliver close to 100 million words/sec meaningful throughput.

\section{Conclusion}
\label{sec:conclusion}

A high performance parallel \wv algorithm in shared and distributed memory systems is proposed. It combines the idea of Hogwild, minibatching and shared negative sampling to convert the level-1 BLAS vector-vector operations to the level-3 BLAS matrix multiply operations. As a result, the proposed algorithm is more hardware-friendly and can efficiently leverage the vector units and multiply-add instruction of modern multi-core and many-core architectures. We also explore different techniques, such as sub-model synchronization and learning rate scheduling, to parallelize the \wv computation across multiple computing nodes. These techniques dramatically reduce network communication and keep the model synchronized effectively when number of nodes increases. We demonstrate the throughput and predictive accuracy of our algorithm comparing to the state-of-the-arts implementations, such the original \wv and BIDMach, on both single node shared memory systems and multi-node distributed systems. We achieve near linear scalability across cores and nodes, and process hundreds of mullions of words per second, the best performance reported so far on the one billion word benchmark.

As for future work, our plans include asynchronous model update similar to parameter sever~\cite{LiAndSmo14}, more efficient sub-model synchronization strategy as well as improving the rate of convergence of the distributed \wv implementation.


\bibliographystyle{IEEEtran}
\bibliography{ref}

\begin{thebibliography}{10}
\providecommand{\url}[1]{#1}
\csname url@samestyle\endcsname
\providecommand{\newblock}{\relax}
\providecommand{\bibinfo}[2]{#2}
\providecommand{\BIBentrySTDinterwordspacing}{\spaceskip=0pt\relax}
\providecommand{\BIBentryALTinterwordstretchfactor}{4}
\providecommand{\BIBentryALTinterwordspacing}{\spaceskip=\fontdimen2\font plus
\BIBentryALTinterwordstretchfactor\fontdimen3\font minus
  \fontdimen4\font\relax}
\providecommand{\BIBforeignlanguage}[2]{{%
\expandafter\ifx\csname l@#1\endcsname\relax
\typeout{** WARNING: IEEEtran.bst: No hyphenation pattern has been}%
\typeout{** loaded for the language `#1'. Using the pattern for}%
\typeout{** the default language instead.}%
\else
\language=\csname l@#1\endcsname
\fi
#2}}
\providecommand{\BIBdecl}{\relax}
\BIBdecl

\bibitem{ManningHinrich99}
C.~D. Manning and H.~Sch\"{u}tze, \emph{Foundations of Statistical Natural
  Language Processing}.\hskip 1em plus 0.5em minus 0.4em\relax Cambridge, MA,
  USA: MIT Press, 1999.

\bibitem{CollobertWeston08}
R.~Collobert and J.~Weston, ``A unified architecture for natural language
  processing: deep neural networks with multitask learning,'' in
  \emph{Proceedings of the 25th international conference on Machine learning},
  2008, pp. 160--167.

\bibitem{GloBorBen11}
X.~Glorot, A.~Bordes, and Y.~Bengio, ``A unified architecture for natural
  language processing: deep neural networks with multitask learning,'' in
  \emph{Proceedings of the 25th international conference on Machine learning},
  2011, pp. 513--520.

\bibitem{Turney13}
P.~D. Turney, ``Distributional semantics beyond words: Supervised learning of
  analogy and paraphrase,'' in \emph{Transactions of the Association for
  Computational Linguistics (TACL)}, 2013, pp. 353--366.

\bibitem{MikSutChe13}
T.~Mikolov, I.~Sutskever, K.~Chen, G.~S. Corrado, and J.~Dean, ``Distributed
  representations of words and phrases and their compositionality,'' in
  \emph{Advances in Neural Information Processing Systems 26}, 2013, pp.
  3111--3119.

\bibitem{ChoMerGul14}
K.~Cho, B.~van Merrienboer, C.~Gulcehre, D.~Bahdanau, F.~Bougares, H.~Schwenk,
  and Y.~Bengio, ``Learning phrase representations using rnn encoder-decoder
  for statistical machine translation,'' in \emph{Proceedings of the 2002
  Conference on Empirical Methods in Natural Language Processing (EMNLP)},
  2014.

\bibitem{WesChoBor15}
J.~Weston, S.~Chopra, and A.~Bordes, ``Memory networks,'' in
  \emph{International Conference on Learning Representations (ICLR)}, 2015.

\bibitem{BlaDemDon02}
L.~S. Blackford, J.~Demmel, J.~Dongarra, I.~Duff, S.~Hammarling, G.~Henry,
  M.~Heroux, L.~Kaufman, A.~Lumsdaine, A.~Petitet, R.~Pozo, K.~Remington, and
  R.~C. Whaley, ``An updated set of basic linear algebra subprograms (blas),''
  \emph{ACM Trans. Mathematical Software}, vol.~28, no.~2, pp. 135--151, 2002.

\bibitem{NiuRecRe11}
F.~Niu, B.~Recht, C.~Re, and S.~J. Wright, ``Hogwild: A lock-free approach to
  parallelizing stochastic gradient descent,'' in \emph{Advances in Neural
  Information Processing Systems}, 2011, pp. 693--701.

\bibitem{CanZhaChe15}
J.~Canny, H.~Zhao, Y.~Chen, B.~Jaros, and J.~Mao, ``Machine learning at the
  limit,'' in \emph{IEEE International Conference on Big Data}, 2015.

\bibitem{DucHazSin11}
J.~Duchi, E.~Hazan, and Y.~Singer, ``Adaptive subgradient methods for online
  learning and stochastic optimization,'' \emph{Journal of Machine Learning
  Research}, vol.~12, pp. 2121--2159, 2011.

\bibitem{Hinton12}
G.~Hinton, ``Lecture 6.5-rmsprop: Divide the gradient by a running average of
  its recent magnitude,'' 2012, cOURSERA: Neural Networks for Machine Learning.

\bibitem{CheMikSch14}
C.~Chelba, T.~Mikolov, M.~Schuster, Q.~Ge, T.~Brants, P.~Koehn, and
  T.~Robinson, ``One billion word benchmark for measuring progress in
  statistical language modeling,'' in \emph{INTERSPEECH}, 2014, pp. 2635--2639.

\bibitem{MillerCharles91}
G.~A. Miller and W.~G. Charles, ``Contextual correlates of semantic
  similarity,'' in \emph{Language and cognitive processes}, 1991.

\bibitem{MikCheCorDea13}
T.~Mikolov, K.~Chen, G.~Corrado, and J.~Dean, ``Efficient estimation of word
  representations in vector space,'' \emph{Proceedings of Workshop at ICLR},
  2013.

\bibitem{ZhangJordan15}
Y.~Zhang and M.~I. Jordan, ``Splash: User-friendly programming interface for
  parallelizing stochastic algorithms,'' \emph{arXiv preprint
  arXiv:1506.07552}, 2015.

\bibitem{FinGabMat02}
L.~Finkelstein, E.~Gabrilovich, Y.~Matias, E.~Rivlin, Z.~Solan, G.~Wolfman, and
  E.~Ruppin, ``Placing search in context: The concept revisited,'' \emph{ACM
  Transactions on Information Systems}, vol.~20, pp. 116--131, 2002.

\bibitem{LiAndSmo14}
M.~Li, D.~Andersen, A.~Smola, J.~Park, A.~Ahmed, V.~Josifovski, J.~Long,
  E.~Shekita, and B.-Y. Su, ``Scaling distributed machine learning with the
  parameter server,'' in \emph{Operating Systems Design and Implementation
  (OSDI)}, 2014.

\end{thebibliography}

\end{document}